\documentclass[12pt]{article}
\usepackage{fancyhdr}

\usepackage{mathrsfs}
\usepackage[T1]{fontenc}
\usepackage{mathpazo}
\usepackage{setspace}
\usepackage{amsfonts}
\usepackage{amssymb}
\usepackage{amsmath}
\usepackage{epsfig}
\usepackage{latexsym}
\usepackage{color}
\usepackage{graphicx}
\usepackage{nicefrac}
\usepackage[latin1]{inputenc}
\usepackage{slashed}
\usepackage{multirow}
\usepackage{comment}
\usepackage{soul}
\usepackage{hyperref}
\usepackage[nosort]{cite}
\usepackage{datetime}

\usepackage{braket}

\usepackage[titletoc]{appendix}

\def\hybrid{\topmargin -20pt    \oddsidemargin 0pt
        \headheight 0pt \headsep 0pt
        \textwidth 6.25in       
        \textheight 9.25in       
        \marginparwidth .875in
        \parskip 5pt plus 1pt   \jot = 1.5ex}

\hybrid

\def\baselinestretch{1.2}

\catcode`\@=11

\def\marginnote#1{}
\newcount\hour
\newcount\minute
\newtoks\amorpm
\hour=\time\divide\hour by60
\minute=\time{\multiply\hour by60 \global\advance\minute by-\hour}
\edef\standardtime{{\ifnum\hour<12 \global\amorpm={am}%
        \else\global\amorpm={pm}\advance\hour by-12 \fi
        \ifnum\hour=0 \hour=12 \fi
        \number\hour:\ifnum\minute<10 0\fi\number\minute\the\amorpm}}
\edef\militarytime{\number\hour:\ifnum\minute<10 0\fi\number\minute}

\def\draftlabel#1{{\@bsphack\if@filesw {\let\thepage\relax
   \xdef\@gtempa{\write\@auxout{\string
      \newlabel{#1}{{\@currentlabel}{\thepage}}}}}\@gtempa
   \if@nobreak \ifvmode\nobreak\fi\fi\fi\@esphack}
        \gdef\@eqnlabel{#1}}
\def\@eqnlabel{}
\def\@vacuum{}
\def\draftmarginnote#1{\marginpar{\raggedright\scriptsize\tt#1}}

\def\draft{\oddsidemargin -.5truein
        \def\@oddfoot{\sl preliminary draft \hfil
        \rm\thepage\hfil\sl\today\quad\militarytime}
        \let\@evenfoot\@oddfoot \overfullrule 3pt
        \let\label=\draftlabel
        \let\marginnote=\draftmarginnote
   \def\@eqnnum{(\theequation)\rlap{\kern\marginparsep\tt\@eqnlabel}%
\global\let\@eqnlabel\@vacuum}  }

\def\preprint{\twocolumn\sloppy\flushbottom\parindent 2em
        \leftmargini 2em\leftmarginv .5em\leftmarginvi .5em
        \oddsidemargin -.5in    \evensidemargin -.5in
        \columnsep .4in \footheight 0pt
        \textwidth 10.in        \topmargin  -.4in
        \headheight 12pt \topskip .4in
        \textheight 6.9in \footskip 0pt
        \def\@oddhead{\thepage\hfil\addtocounter{page}{1}\thepage}
        \let\@evenhead\@oddhead \def\@oddfoot{} \def\@evenfoot{} }

\def\numberbysection{\@addtoreset{equation}{section}
        \def\theequation{\thesection.\arabic{equation}}}

\def\underline#1{\relax\ifmmode\@@underline#1\else
        $\@@underline{\hbox{#1}}$\relax\fi}

\def\titlepage{\@restonecolfalse\if@twocolumn\@restonecoltrue\onecolumn
     \else \newpage \fi \thispagestyle{empty}\c@page\z@
        \def\thefootnote{\fnsymbol{footnote}} }

\def\endtitlepage{\if@restonecol\twocolumn \else \newpage \fi
        \def\thefootnote{\arabic{footnote}}
        \setcounter{footnote}{0}}  

\catcode`@=12
\relax

\def\figcap{\section*{Figure Captions\markboth
        {FIGURECAPTIONS}{FIGURECAPTIONS}}\list
        {Figure \arabic{enumi}:\hfill}{\settowidth\labelwidth{Figure
999:}
        \leftmargin\labelwidth
        \advance\leftmargin\labelsep\usecounter{enumi}}}
 \relax
\def\tablecap{\section*{Table Captions\markboth
        {TABLECAPTIONS}{TABLECAPTIONS}}\list
        {Table \arabic{enumi}:\hfill}{\settowidth\labelwidth{Table
999:}
        \leftmargin\labelwidth
        \advance\leftmargin\labelsep\usecounter{enumi}}}
 \relax
\def\reflist{\section*{References\markboth
        {REFLIST}{REFLIST}}\list
        {[\arabic{enumi}]\hfill}{\settowidth\labelwidth{[999]}
        \leftmargin\labelwidth
        \advance\leftmargin\labelsep\usecounter{enumi}}}
 \relax
\makeatletter
\newcounter{pubctr}
\def\publist{\@ifnextchar[{\@publist}{\@@publist}}
\def\@publist[#1]{\list
        {[\arabic{pubctr}]\hfill}{\settowidth\labelwidth{[999]}
        \leftmargin\labelwidth
        \advance\leftmargin\labelsep
        \@nmbrlisttrue\def\@listctr{pubctr}
        \setcounter{pubctr}{#1}\addtocounter{pubctr}{-1}}}
\def\@@publist{\list
        {[\arabic{pubctr}]\hfill}{\settowidth\labelwidth{[999]}
        \leftmargin\labelwidth
        \advance\leftmargin\labelsep
        \@nmbrlisttrue\def\@listctr{pubctr}}}
 \relax
\makeatother
\newskip\humongous \humongous=0pt plus 1000pt minus 1000pt

\newif\ifdtup

\relax

\def\be{\begin{equation}}
\def\ee{\end{equation}}
\def\ba{\begin{eqnarray}}
\def\ea{\end{eqnarray}}

\def\del{\partial}

\def\r{\rho}

\def\b{\beta}

\def\g{\gamma}

\def\d{\delta}

\def\e{\epsilon}

\def\m{\mu}

\def\l{\lambda}

\def\s{\sigma}

\def\cL{{\cal L}}

\def\no{\noindent}

\def\qq{\qquad}

\def\IR{\relax{\rm I\kern-.18em R}}

\def \z { {\bar z} }
\def \w { {\bar w} }

\def \ha {{1\over 2}}

\def \ov {\over}

\def\IR{\relax{\rm I\kern-.18em R}}
\def\IL{\relax{\rm I\kern-.18em L}}

\def\inv{^{\raise.15ex\hbox{${\scriptscriptstyle -}$}\kern-.05em 1}}

\def\cL{{\cal L}}

\def\Tr{{\rm Tr}}

\begin{document}

\renewcommand{\theequation}{\thesection.\arabic{equation}}
\csname @addtoreset\endcsname{equation}{section}

\newcommand{\beq}{\begin{equation}}
\newcommand{\eeq}[1]{\label{#1}\end{equation}}
\newcommand{\ber}{\begin{equation}}
\newcommand{\eer}[1]{\label{#1}\end{equation}}
\newcommand{\eqn}[1]{(\ref{#1})}
\begin{titlepage}
\begin{center}


${}$
\vskip .2 in

\vskip .4cm

{\large\bf
Field theory and $\l$-deformations: Expanding around the identity
}

\vskip 0.4in

{\bf George Georgiou}\hskip .2 cm and \hskip .15 cm {\bf Konstantinos Sfetsos}
\vskip 0.16in

 {\em
Department of Nuclear and Particle Physics,\\
Faculty of Physics, National and Kapodistrian University of Athens,\\
Athens 15784, Greece\\
}

\vskip 0.12in

{\footnotesize \texttt {ggeo, ksfetsos@phys.uoa.gr}}


\vskip .5in
\end{center}

\centerline{\bf Abstract}

\no
We explore the structure of the $\l$-deformed $\s$-model action by setting up a perturbative expansion around the free field point corresponding to the identity group element.
We include all field interaction terms up to sixth order.
We compute the two- and three-point functions of current and primary field operators, their anomalous dimensions as well as the $\beta$-function for the $\l$-parameter.
Our results are in complete agreement with those obtained previously using gravitational and/or CFT perturbative methods in conjunction  with the non-perturbative symmetry, as well as with those obtained using methods exploiting the geometry defined in the space of couplings.
The advantage of this approach is that all deformation effects are already encoded in the couplings of the interaction vertices and in the $\l$-dressed operators.

\vskip .4in
\noindent
\end{titlepage}
\vfill
\eject

\newpage

\tableofcontents

\noindent

\def\baselinestretch{1.2}
\baselineskip 20 pt
\noindent


\setcounter{equation}{0}

\section{Introduction }

Integrability plays a  central r\^ole in modern theoretical physics. It is present in QCD at the high energy regime \cite{Faddeev:1994zg}, 
it emerges in several realizations of the gauge/gravity correspondence \cite{Minahan:2002ve,Bena:2003wd} and it is also employed to model condensed matter systems 
by the use of integrable spin chains. Practically, it provides not only the information that a theory is solvable for any value of the coupling constant, but in addition tools for solving it are provided. 

A whole class of integrable two-dimensional field theories having an explicit action representation, were systematically constructed 
\cite{Sfetsos:2013wia,Georgiou:2016urf,Georgiou:2017jfi,Sfetsos:2017sep,Georgiou:2018hpd,Georgiou:2018gpe,Driezen:2019ykp}
and studied \cite{Sfetsos:2014lla,Georgiou:2015nka,Georgiou:2016iom,Georgiou:2016zyo,Georgiou:2017oly,Itsios:2014lca,Georgiou:2017aei}.
These theories go under the name of $\l$-deformations. The most straightforward way to construct them is by a specific 
gauging procedure initiated in \cite{Sfetsos:2013wia}. They are parametrized by one or several matrices, for small entries of which the models are nothing but a WZW model \cite{Witten:1983ar} (or several ones) perturbed by current bilinears. 
In that respect, these models represent the effective actions of WZW current algebra conformal field theories (CFTs) perturbed by current bilinears. As such they take into account all perturbative effects in the perturbation/deformation parameters. 
The simplest of these models is the single $\l$-deformed model \cite{Sfetsos:2013wia} (for the $SU(2)$ group case this model has appeared before in \cite{Balog:1993es}). In \cite{Georgiou:2016iom,Georgiou:2015nka},  two- and three-point correlation functions of single currents and primary field operators were calculated as exact functions of the deformation parameter $\l$.
This was achieved by using low order perturbation theory around the conformal point assisted by a certain non-perturbative  symmetry in the space of couplings which these models generically exhibit. More recently, the exact in $\l$ anomalous dimensions of certain composite current operators were calculated in \cite{Georgiou:2019jcf}. To establish that, a novel method combining geometrical data in the space of couplings and the all-loop effective action was invented.
This method allows in principle the calculation of the anomalous dimensions of composite operators built from an arbitrary number of currents, although the huge operator mixing problem makes the computations for general operators cumbersome.

The aim of the present work is to present yet another method to compute all the aforementioned observables, as exact functions of $\l$, which in certain aspects is simpler and advantageous. 
The main idea is that instead of using perturbation theory around the conformal point, 
to use the free field theory point as a reference. 
It turns out that this corresponds to setting up a large $k$ perturbative expansion of the effective action around the free field point associated with the identity group element. We include all field interaction terms up to sixth order. \footnote{A similar expansion was recently considered in \cite{Hoare:2018jim} aiming at studying the properties of massless tree level S-matrices for two-dimensional  $\sigma$-models.}The resulting action will be a two-dimensional quantum field theory with a canonical kinetic term, but in principle one may keep an infinite number of interaction terms involving successively more and more fields. 
One of the virtues of our method is that it incorporates automatically  all the dependence in $\l$. In fact all the vertices are invariant under the aforementioned non-pertubative symmetry of the model.  In contradistinction with the analogous 
perturbative calculations around the conformal point, this non-perturbative symmetry does not have to be imposed in order to obtain the exact in $\l$ form of the correlators but is already built in the formalism.
As a result, all the results obtained by using this action will inherit this symmetry and will be exact in $\l$. A second advantage of our approach is that it can be considered as the basis for systematically performing the perturbative expansion in powers of ${1/k}$. 

The plan of the paper is as follows: In section 2, we will perform the perturbative expansion of the action around the free field
point. In section 3, we first find the free field expansion of  the fundamental currents of the theory. Subsequently, we use these and the above action in order to calculate three-point functions involving purely chiral or anti-chiral currents
and well as mixed ones,  as exact functions of the deformation parameter $\l$. In the same section, we will also calculate the anomalous dimensions of these fundamental currents, as well as those of primary fields.
In section 4, we calculate the, exact in $\l$, $\b$-function of the model by considering the renormalization of the cubic vertex. For consistency, we present the non-trivial check that the same expression for $\b$-function can be obtained from the renormalization of the quatric coupling.
All the results are in complete agreement with the expressions obtained by the other methods briefly mentioned above.
Finally, in section 5 we will present our conclusions and future directions of this work.

\section{Expansion around the free point}

Our starting point will be he $\l$-deformed action for an element $g$ of a semi-simple group $G$ 
given by \cite{Sfetsos:2013wia}
\be
\label{djkg11}
S_{k,\l}(g)= S_k(g) + {k\ov \pi} \int d^2\s\  J_+ (\l^{-1}\mathbb{1} - D^T)^{-1} J_-\ ,
\ee
where the WZW action is
\be
S_k(g) = - {k\ov 2\pi} \int d^2\s\  \Tr(g^{-1}\del_+ g g^{-1}\del_-g) + {k\ov 12\pi} \int (g^{-1} dg)^3
\ee
and
\be
\label{jjd}
J_+ =- i\, \del_+ g g^{-1} \ ,\qquad J_- = - i\, g^{-1} \del_- g\ ,\qquad D_{ab}=\Tr(t^a g t^b g^{-1})\ .
\ee
The representation matrices are denoted by $t^a$ and normalized to unity. In addition,  they obey the relation $[t_a,t_b]=i f_{abc} t_c$ for some set of real structure constants.
The above action can be rewritten as
\be
\label{sdddg}
S_{k,\l}(g)= {k\ov 2\pi} \int d^2\s\ J_+\, { D+\l\mathbb{1}\ov \mathbb{1} - \l D^T}\, J_-\,
+\, {k\ov 12\pi} \int (g^{-1} dg)^3\ .
\ee
In what follows, it will be particularly useful to parametrize the group element $g\in G$ in terms of  normal coordinates as
\be
\label{ggrouu}
g=e^{i t_a x^a}\ ,
\ee
leading to the following expressions for the matrix in the adjoint represantion and for the currents
\be
\begin{split}
&
D =e^f = \mathbb{1}+ f +{ f^2\ov 2} + \cdots \ ,
\\
&
J_- = { \mathbb{1}-e^{-f} \ov f}\del_- x =  \Big(\mathbb{1} -{f\ov 2} +{f^2\ov 6} + \cdots\Big)\del_- x \ ,
\\
&
J_+  = {e^f- \mathbb{1}\ov f}\del_+x =  \Big(\mathbb{1} +{f\ov 2} +{f^2\ov 6} + \cdots\Big)\del_+ x \ ,
\end{split}
\ee
We have used the definition $f_{ab} = f_{abc}x^c$
and we have expanded the above quantities for small $x$'s around the identity.

The next step is to organize the expansion around the identity as a large $k$ expansion.
We let for the nornal coordinates the rescaling
\be
\label{xphii}
x^a = \sqrt{1-\l\ov 1+\l}\, {\phi^a\ov \sqrt{k}}\ ,
\ee
which as we will see, allows the new fields $\phi^a$'s to have a canonically normalized kinetic term in the action.

\no
It turns out that we will need the action \eqn{sdddg} up to ${\cal O}(1/k^{2})$ in the
large $k$ expansion. Using the above coordinates, the first term of the action \eqn{sdddg}
gives to the specified order in the large-$k$ expansion the action
\be
\begin{split}
& S_{k,\l}^{(2)} = {1\ov 2\pi} \int d^2\s\, \Big(\del_+\phi^a\del_-\phi^a
+ {g_3^{(2)}\ov \sqrt{k}}\, f_{ab}\,\del_+\phi^a\del_-\phi^b
\\
&\qq\ + {g_4^{(2)}\ov k} \, f^2_{ab}\, \del_+\phi^a\del_-\phi^b
+  {g_5^{(2)}\ov k^{3/2}} \, f^3_{ab}\, \del_+\phi^a\del_-\phi^b \Big)
+ {g_6^{(2)}\ov k^2} \, f^4_{ab}\, \del_+\phi^a\del_-\phi^b\Big)
+\cdots\ ,
\end{split}
\ee
where
\be
\begin{split}
& g_3^{(2)}=  - {2\l\ov (1-\l)^{1/2} (1+\l)^{3/2}}\ ,\qq
g_4^{(2)}= {1+10\l+\l^2\ov 12(1-\l^2)}\ ,\qq
\\
&
g_5^{(2)}=  - {\l\ov 2(1-\l)^{3/2} (1+\l)^{1/2}}\ ,\qq
g_6^{(2)}= {1 + 56 \l + 246 \l^2 + 56 \l^3 + \l^4)\ov 360 (1 - \l^2)^2}\ .
\end{split}
\ee
Also we have redefined  $f_{ab}$ in terms of the field $\phi$ 
\be
f_{ab} = f_{abc}\phi^c\ ,
\ee
a definition which will be used in the rest of the paper.
In this expansion, we have kept the free part properly normalized as well as interactions up to the 
sixth order.

Now we turn to the expansion of the topological term, that is  the second term in \eqn{sdddg}.
The expansion of this term can be obtained by first considering the expansion of corresponding field strength
\be
H_0 = -{k\ov 6} f_{abc} L^a \wedge L^b \wedge L^c\ , \qq L^a =- i\, \Tr(t^a g^{-1} dg)\ .
\ee
We need to expand $H_0$ to ${\cal O}(1/k^2)$ and then read off the corresponding
two-form antisymmetric tensor $B_0$ and its contribution to the action.
After certain algebraic manipulations using mainly the Jacobi identity for the structure
constants, we find that
\be
\begin{split}
& H_0= -{1\ov 6\sqrt{k}} \Big({1-\l\ov 1+\l}\Big)^{3/2} f_{abc} d\phi^a\wedge d\phi^b \wedge d\phi^c
\\
& \qq \ + {1\ov 24 k^{3/2}}  \Big({1-\l\ov 1+\l}\Big)^{5/2}
f_{abc} f_{ad} f_{be} d\phi^d\wedge d\phi^e \wedge d\phi^c  + \cdots \ .
\end{split}
\ee
Note that there are no ${\cal  O}(1/k)$ and  ${\cal  O}(1/k^2)$ terms.
Using the above expression we may read off the antisymmetric tensor field from the relation
$H_0=dB_0$.   We emphasize that this is a local expression. Then, we find out that the topological term contributes to the action as
\be
S_{k,\l}^{(3)} = {1\ov 2\pi}  \int d^2\s\,
\Big({g_3^{(3)}\ov \sqrt{k}} f_{ab}\, \del_+\phi^a\del_-\phi^b
+  {g_5^{(3)}\ov k^{3/2}} f^3_{ab}\, \del_+\phi^a\del_-\phi^b\Big)+ \cdots \ ,
\ee
where
\be
g_3^{(3)} = -{1\ov 3}\Big({1-\l\ov 1+\l}\Big)^{3/2} \ ,\qq
g_5^{(3)} = -{1\ov 60 }\Big({1-\l\ov 1+\l}\Big)^{5/2}\ .
\ee
As a remark we mention that in the expansion of the quadratic action all powers of the matrix $f$
appear, whereas for the cubic part of the action  only the odd ones do so.

Putting everything altogether \eqn{sdddg} assumes the expansion
\be
\boxed{
\begin{split}
\label{sklinf}
& S_{k,\l} = {1\ov 2\pi} \int d^2\s\, \Big(\del_+\phi^a\del_-\phi^a +{g_3\ov \sqrt{k}} f_{ab}\del_+\phi^a\del_-\phi^b
 \\
 &\qq\qq\ + {g_4\ov k} f^2_{ab}\del_+\phi^a\del_-\phi^b
  + {g_5\ov k^{3/2}} f^3_{ab}\del_+\phi^a\del_-\phi^b
  +  {g_6\ov k^2} f^4_{ab}\del_+\phi^a\del_-\phi^b
   \Big)+\cdots\ ,
\end{split}
}
\ee
where the couplings are $g_i = g_i^{(2)}+ g_i^{(3)}$, $i=3,\dots , 6$. They assume the form
\be
\label{g34}
\begin{split}
& g_3(\l)= -{1\ov 3} {1+4\l+\l^2\ov (1-\l)^{1/2}(1+\l)^{3/2}}\ ,\qquad g_4(\l)= {1\ov 12}{1+10\l+\l^2\ov(1-\l^2)}\ ,
\\
& g_5(\l)= - {1\ov 60} {1+26\l+66\l^2 + 26 \l^3 +\l^4\ov (1-\l)^{3/2}(1+\l)^{5/2}} \ ,
\\
&
g_6(\l)= {1\ov 360} {1 + 56 \l + 246 \l^2 + 56 \l^3 + \l^4)\ov(1 - \l^2)^2}\ .
\end{split}
\ee
Note the symmetry of \eqn{sklinf} under
\be
\label{invv}
\l\to {1\ov \l} \ , \qq k\to -k\ , \qq \phi^a\to -\phi^a\ .
\ee
This originates form the same symmetry of \eqn{sklinf} discovered in the present context in \cite{Itsios:2014lca}
and earlier using path integral arguments in \cite{Kutasov:1989aw}, where the group element
inversion corresponds to flipping the sign of $\phi^a$.
Note also  that, since there are square roots involved in
the definition of the coupling $g_3$, we should refine the way the aforementioned symmetry acts.
In particular, we have that $k\to e^{i\pi} k $ and $1-\l\to e^{i\pi} (\l^{-1}-1)$ implying, for instance, that $\sqrt{k (1-\l)}\to -\sqrt{k (1-\l)}\,\l^{-1/2}$.

\no
The reader might wonder for the reason we have kept in our expansion terms as high as of those of order six in the
fields. As we shall see, in determining the three- and two-point functions for currents to ${\cal O}(1/\sqrt{k})$ and ${\cal O}(1/k)$, respectively, only $g_3$ is necessary. However, for determining the $\beta$-function from the renormalization of the cubic interaction to ${\cal O}(1/k)$ all three $g_3,g_4$ and $g_5$ should be kept. Finally, to verify that the same
$\b$-function follows from the renormalization of the quartic interaction the coupling $g_6$ is
also necessary.

\subsection{Computational QFT conventions}

We would like to set up a perturbative expansion around the free theory. Passing to the
Euclidean regime we have the following basic propagators
\be
\langle \phi^a(z,\z) \phi^b(w,\w)\rangle = -\d^{ab}\ln |z-w|^2\ ,
\ee
which is consistent with our normalizations.
We will use the notation
\be
\label{noott}
J^a(z) = \del \phi^a(z,\z)\ ,\qq \bar J^a(z) = \bar\del \phi^a(z,\z)\ .
\ee
Then, we have that
\be
\label{opeee}
\begin{split}
&\langle J^a(z) \phi^b(w,\w)\rangle = -{\d^{ab}\ov z-w}\  , \qq \langle J^a(z) J^b(w) \rangle= -{\d^{ab}\ov (z-w)^2}\ ,
\\
&\langle \phi^a(z,\z) f^{bc}(w,\w)\rangle =-f_{abc}\ln |z-w|^2\ ,
 \\
 &\langle f^{ab}(z,\z)f^{cd}(w,\w)\rangle =-f_{abe}f_{cde}\ln |z-w|^2 \ ,
\\
&
\langle J^a(z) f^{bc}(w,\w)\rangle =-{f_{abc}\ov z-w}\ .
\end{split}
\ee
Finally, note the propagator
\be
\langle J^a(z)\bar J^b(w)\rangle =C\, \d^{ab}\d^{(2)}(z-w)\ ,
\ee
which couples the holomorphic and anti-holomorphic sectors. 
The constant $C$ is\footnote{ Note that we differ by a factor of $1/2$ from other conventions,  
e.g. eq. (2.5.8) of \cite{Pol}. This apparent disagreement comes from the fact that in our conventions $\d^{(2)}(z)=\d(x)\d(y):=\d^{(2)}({\bf x})$, while in  \cite{Pol}
$\d^{(2)}(z)=\ha \d^{(2)}({\bf x})$  (see eq. (2.1.8) in that reference and the line below it).
Accordingly, in our conventions $z=x+iy$.  In addition, the measure of integration in the 
Euclidean regime is $d^2z = dxdy$.
}
 \be
\label{CCC}
 C=\pi \ ,
 \ee
 which is proven by taking the derivative of the first propagator in the first line in \eqn{opeee}
 with respect to $\bar w$ and subsequently using
that $\displaystyle \del_z {1\ov \z} = \del_\z {1\ov z}= \pi \d^{(2)}(z)$.\footnote{
These are  also consistent with the relation 
$(\del_x^2+\del_y^2) \ln \r =  2\pi \d^{(2)}({\bf x})$,
proven using Stoke's theorem in two-dimensions, as well as with the previous footnote.
}

\section{Correlation functions and anomalous dimensions}

\no
We are interested in the correlation function of  the currents $J_\pm$. However, these after the
deformation no longer have the form \eqn{jjd}, but they are instead $\l$-dressed.
The dressed currents has been identified in\cite{Georgiou:2016iom} with the on-shell values of the gauge fields arising from
integrating them out in the process of constructing the $\l$-deformed action \eqn{djkg11}.
These have the following expressions
\be
\label{apam1}
A_+= i \big(\l^{-1}\mathbb{1} - D)^{-1} J_+\ , \qq A_- =-  i\big(\l^{-1}\mathbb{1} - D^T)^{-1} J_-\ ,
\ee
They are not invariant under  the transformation \eqn{invv}, but they can be made so
by multiplying with an appropriate $\l$-dependent factor. Obviously this will not affect their
anomalous dimensions.
One should expand them as we did for the action. For a well defined large
$k$-expansion we should first multiply them with an appropriate constant. This is chosen so that
 such the leading order term is $ \pm\del_\pm \phi$.
 Indeed, multiplying $A_\pm $  by
$i \sqrt{k} (1-\l^{-1})$ and denoting the result by ${\cal J}_\pm$ we have that
\be
\boxed{
\label{jjdre}
\begin{split}
&{\cal J}_+ = \Big(\mathbb{1} +  {h_1\ov  \sqrt{k}} f  + {h_2 \ov k} f^2+\cdots
\Big)\del_+\phi\ ,
\\
&
{\cal J}_- = -\Big(\mathbb{1} -  {h_1\ov  \sqrt{k}} f  + {h_2 \ov k} f^2+\cdots
\Big)\del_-\phi\ ,
\end{split}
}
\ee
where
\be
\label{g12}
h_1(\l)= \ha \sqrt{1+\l\ov 1-\l}\ ,\qq h_2(\l)= {1+4\l+\l^2\ov 6(1-\l^2)}\ .
\ee
We will see that for computing their two- and three-point correlation functions and their anomalous dimensions up to
${\cal O}(1/k)$ only the coefficient $h_1$ will play a role. Note also that the normalization of the
fields is such that at the conformal point for $\l=0$, the coefficient of the non-Abelian term
in their operator product expansions is proportional to $f^{abc}/\sqrt{k}$. The precise coefficient is $-3$ and not $1$ as is probably expected. The reason for the discrepancy is that in the $\l=0$ limit one still has to take into account the interaction terms in \eqn{sklinf}. Hence, \eqn{jjdre} should not
be thought of as a bosonization formula.

\no
In the Euclidean regime we will use $\cal J$ and $\bar {\cal  J}$ in place of
$\cal J_+$ and $ \cal  J_-$, respectively. 

\subsection{The three-point function for currents}

The easiest correlators to compute involving only currents are three-point ones.
The leading result is of ${\cal O}(1/\sqrt{k})$. For the purely chiral correlator and
employing a self-explanatory notation for the $\l$-dependent correlators, we have that
\be
\langle {\cal J}^a(x_1,\bar x_1) {\cal J}^b(x_2,\bar x_2)  {\cal J}^c(x_3,\bar x_3)\rangle_\l
=\langle {\cal J}^a(x_1,\bar x_1) {\cal J}^b(x_2,\bar x_2)  {\cal J}^c(x_3,\bar x_3)
e^{-S_{\rm int}}\rangle\ ,
\ee
where the interaction terms in the Euclidean regime can be read from \eqn{sklinf} to be
\be
S_{\rm int} =   {g_3\ov 2\pi \sqrt{k}} \int d^2z\, J^af_{ab}\bar J^b
+  {g_4\ov 2\pi k} \int d^2z\, J^af^2_{ab}\bar J^b +  {g_5\ov 2\pi k^{3/2}} \int d^2z\, J^af^3_{ab}\bar J^b\ .
\ee
Then, expanding the exponential and keeping terms up to ${\cal O}(1/\sqrt{k})$ we have that
\be
\begin{split}
&\langle {\cal J}^a(x_1,\bar x_1) {\cal J}^b(x_2,\bar x_2)  {\cal J}^c(x_3,\bar x_3)\rangle_\l
\\
&
\qquad\ ={h_1\ov \sqrt{k}} \langle f^{aa_1}(x_1,\bar x_1) J^{a_1}(x_1) J^b(x_2) J^c(x_3) \rangle
+ \big[{\rm cyclic\ in}\ (x_1,a), (x_2,b), (x_3,c)\big]
\\
&  \qquad\quad\  -{g_3\ov 2\pi \sqrt{k}}\int d^2z\
\langle J^a(x_1) J^b(x_2) J^c(x_3) J^{a_1}(z) f^{a_1b_1}(z,\z)
\bar J^{b_1}(\z) \rangle
\\
& \qq\ ={1\ov \sqrt{k}} \Big(h_1  + { C\ov 2\pi}g_3\Big) \langle f^{aa_1}(x_1,\bar x_1) J^{a_1}(x_1) J^b(x_2) J^c(x_3)\rangle
\\
&
\qq\qq\qq\qq\qq + \big[{\rm cyclic\ in}\ (x_1,a), (x_2,b), (x_3,c)\big] \ .
\end{split}
\ee
Explicitly, the necessary four-point function is given by
\be
\label{EGG1}
 \langle f^{aa_1}(x_1,\bar x_1) J^{a_1}(x_1) J^b(x_2) J^c(x_3) \rangle=
 {f_{abc}\ov x_{12} x_{13}}\bigg({1\ov x_{13}}-{1\ov x_{12}}\bigg)\ ,
\ee
finding finally that
\be
\langle{\cal J}^a(x_1,\bar x_1) {\cal J}^b(x_2,\bar x_2)   {\cal J}^c(x_3,\bar x_3)\rangle_\l
 = {1\ov \sqrt{k}}\, \Big(3 h_1  + {3 C\ov 2\pi}g_3\Big)\,
 {f_{abc}\ov x_{12} x_{13} x_{23}}\ .
\ee
Using for $C$ the value in \eqn{CCC} and also \eqn{g34} and \eqn{g12} we obtain that
\be
\boxed{
\langle{\cal J}^a(x_1,\bar x_1) {\cal J}^b(x_2,\bar x_2)   {\cal J}^c(x_3,\bar x_3)\rangle_\l
 = {1\ov \sqrt{k}}\,  {1+\l+\l^2\ov (1 - \l)^{1/2} (1 + \l)^{3/2}}\,
{f_{abc}\ov x_{12} x_{13} x_{23}}}\ .
\ee
This result agrees with the one  computed before using conformal perturbation theory and the
non-perturbative symmetry in the coupling space \eqn{invv} (see eq. (3.29) of
\cite{Georgiou:2016iom}).

\no
Similarly, for the mixed chirality correlator and keeping only those terms that potentially contribute to the correlator
up to ${\cal O}(1/\sqrt{k})$, we have that
\be \label{corr1}
\begin{split}
&\langle {\cal J}^a(x_1,\bar x_1) {\cal J}^b(x_2,\bar x_2)
\bar{\cal J}^c(x_3,\bar x_3)\rangle_\l
=\langle {\cal J}^a(x_1,\bar x_1) {\cal J}^b(x_2,\bar x_2)
\bar{\cal J}^c(x_3,\bar x_3) e^{-S_{\rm int}}\rangle
\\
&\quad =-
{h_1\ov \sqrt{k}} \langle f^{aa_1}(x_1,\bar x_1) J^{a_1}(x_1) J^b(x_2) \bar J^c(\bar x_3) \rangle
+\big[ (x_1,a) \leftrightarrow (x_2,b)\big]
\\
 & \qq\  +{g_3\ov 2\pi \sqrt{k}}\int d^2z\
\langle J^a(x_1) J^b(x_2) \bar J^c(\bar x_3) J^{a_1}(z) f^{a_1b_1}(z,\z) \bar J^{b_1}(\z) \rangle
\\
&
 \quad = -
 {1\ov \sqrt{k}} \Big(h_1  + { C\ov 2\pi}g_3\Big) \langle f^{aa_1}(x_1,\bar x_1) J^{a_1}(x_1) J^b(x_2) \bar J^c(\bar x_3) \rangle +\big[ (x_1,a) \leftrightarrow (x_2,b)\big]
 \\
 &\qq\ + {g_3\ov 2\pi \sqrt{k}}\int d^2z\ {1\ov (\z-\bar x_3)^2}
 \langle f^{ca_1}(z,\z) J^{a_1}(z) J^a(x_1)  J^b( x_2) \rangle \ .
\end{split}
\ee
In the fourth line of \eqref{corr1} we have  the correlator
\be\label{EGG0}
 \langle f^{aa_1}(x_1,\bar x_1) J^{a_1}(x_1) J^b(x_2) \bar J^c(\bar x_3) \rangle=
- {f_{abc}\ov x_{12}^2\bar x_{13} } - C {f_{abc} \ov x_{12}} \d^{(2)}(x_{12}) \ .
\ee
The last term will be ignored since it is a contact term of external points.
To evaluate the integral in the last line of  \eqref{corr1} we have used \eqn{EGG0} and the complex conjugate of the last integral in \eqn{B2}.
Putting everything together we find that
\be
\langle {\cal J}^a(x_1,\bar x_1) {\cal J}^b(x_2,\bar x_2) \bar{\cal J}^c(x_3,\bar x_3)\rangle_\l
 = -{1\ov \sqrt{k}}\, \Big( h_1 +{2\pi + C\ov 2\pi}g_3\Big)\,
 {f_{abc} \bar x_{12}\ov x_{12}^2 \bar x_{13} \bar x_{23}}\ .
\ee
Substituting $\pi$ for $C$ and the expressions for the couplings $h_1$ and $g_3$ one gets the final result
\be
\boxed{
\langle {\cal J}^a(x_1,\bar x_1) {\cal J}^b(x_2,\bar x_2) \bar{\cal J}^c(x_3,\bar x_3)
\rangle_\l
 = {1\ov \sqrt{k}}\, {\l\ov (1 - \l)^{1/2} (1 + \l)^{3/2}} \,
 {f_{abc} \bar x_{12}\ov x_{12}^2 \bar x_{13} \bar x_{23}}
 }\ .
\ee
Again, the result agrees with the one computed in \cite{Georgiou:2016iom} using conformal perturbation theory and the non-perturbative symmetry in the coupling space \eqn{invv}
(see eq. (3.33) of \cite{Georgiou:2016iom}).
Needless to say, it was an essential part of the computation that the $\l$-dressed currents were used. Had we used the bare currents the correct result would not have been obtained. 

\subsection{The single current anomalous dimension}

In order to obtain the anomalous dimension of the currents we evaluate the two-point function
\be
\langle {\cal J}^a(x_1,\bar x_1) {\cal J}^b(x_2,\bar x_2)\rangle_\l =
\langle {\cal J}^a(x_1,\bar x_1) {\cal J}^b(x_2,\bar x_2)
e^{-S_{\rm int}}\rangle\ .
\ee
Since we are after the leading term of the anomalous dimension which is of order ${\cal O}(1/k)$ we will expand the exponential keeping terms up to the same order.
It can be easily seen that  ${\cal O}(1/\sqrt{k})$ contribution vanishes as it contain only terms with an odd number of fields.
Then the result for the two-point correlator up to ${\cal O}(1/k)$ reads
\be
\label{233}
\langle {\cal J}^a(x_1,\bar x_1) {\cal J}^b(x_2,\bar x_2)\rangle_\l =
-{\d^{ab}\ov x_{12}^2} -  {1 \ov k}\Big(h_1^2 I_1^{ab}
-\ha {g_3^2\ov 4\pi^2} I_2^{ab} - {h_1g_3\ov 2\pi} I_3^{ab}\Big)\ ,
\ee
where we have defined
\be
\label{I123}
\begin{split}
& I_1^{ab} = \langle J^c(x_1) f^{ca}(x_1,\bar x_1) f^{bd}(x_2,\bar x_2) J^d(x_2)\rangle\ ,
\\
&
 I_3^{ab} = \int d^2 z\ \langle J^c(x_1)
f^{ca}(x_1,\bar x_1) J^b(x_2) J^{a_1}(z) f^{a_1 a_2}(z,\bar z) \bar J^{a_2}(\z) \rangle
 + (x_1,a)\leftrightarrow (x_2,b)\ ,
\\
& I_2^{ab} = \int d^2z_1 d^2 z_2\
\langle J^a(x_1) J^b(x_2) J^{a_1}(z_1) f^{a_1a_2}(z_1,\bar z_1)\bar J^{a_2}(\z_1)
\\
&\qq\qq\qq\qq\qq\qq\qq \times J^{b_1}(z_2) f^{b_1 b_2}(z_2,\bar z_2)\bar J^{b_2}(\z_2)\rangle\ .
\end{split}
\ee
By introducing a short distance cut-off $\e$ one can  easily compute the first integral to be
\be
 I_1^{ab}  = -{c_G\d^{ab}\ov x_{12}^2} + {c_G\d_{ab}\ov x_{12}^2}\, 
 \ln{\e^2\ov |x_{12}|^2}\ .
\ee

\no
The integral in the second line of \eqn{I123} is evaluated to give
\be
\begin{split}
I_3^{ab} = & C \int d^2z\ \d^{(2)}(z-x_1)
\langle f^{ca}(x_1,\bar x_1) J^b(x_2) J^{a_1}(z) f^{a_1 c}(z,\z)\rangle
\\
& + C\ \langle J^c(x_1) f^{ca}(x_1,\bar x_1) J^{a_1}(x_2) f^{a_1b}(x_2,\bar x_2)\rangle
\\
& -\int d^2 z\ {f^{caa_2}\ov \z-\bar x_1}
\langle J^c(x_1) J^{b}(x_2) J^{a_1} (z) f^{a_1a_2}(z,\z)\rangle
+  [(x_1,a)\leftrightarrow (x_2,b)]\ .
\end{split}
\ee
Using the first and second integral in \eqn{B2} one  finds that
\be
I_3^{ab}=  2 (C-\pi) {c_G \d^{ab}\ov x_{12}^2} - 2 (2\pi +C) {c_G \d^{ab}\ov x_{12}^2}\,
\ln{\e^2\ov |x_{12}|^2}\ .
\ee
Using $C=\pi$  we obtain the result
\be
\label{i3ab}
I_3^{ab}=  - 6\pi\, {c_G \d^{ab}\ov x_{12}^2}\,
\ln{\e^2\ov |x_{12}|^2}\ .
\ee
Finally, we should consider the integral $I_2^{ab}$. This requires a more involved computation the details of  which are presented in the appendix. 
The end result, reads
\be
\label{i2ab}
I_{2}^{ab}=  - 18\pi^2\, {c_G \d^{ab}\ov x_{12}^2}\,
\ln{\e^2\ov |x_{12}|^2}\ .
\ee
Consequently, \eqn{233} gives
\be
\label{233f}
 \langle {\cal J}^a(x_1,\bar x_1) {\cal J}^b(x_2,\bar x_2)\rangle_\l =
-{\d^{ab}\ov x_{12}^2}\Bigg( 1 - {c_G\ov k} h_1^2+  {c_G \ov k}\Big(h_1
+{3\ov 2} g_3\Big)^2 \ln{\e^2\ov |x_{12}|^2}\Bigg)\, ,
\ee
Inserting in the last equation the expressions for $g_3$ and  $h_1$ that can be found in \eqn{g34} and \eqn{g12} respectively, one can find that up to ${\cal O}(1/k)$ the current anomalous dimension
is given by
\be
\label{singlJ}
\boxed{\g = {c_G\ov k} {\l^2\ov (1 - \l) (1 + \l)^3}}\ .
\ee
This is the same result found using either CFT perturbation theory in combination with the symmetry \eqn{invv} in coupling space \cite{Georgiou:2015nka} or the effective action and the geometry in coupling space in \cite{Georgiou:2019jcf}. Obviously, we would have reached the same
conclusion for the anomalous dimension if we had considered the two-point correlator
for $\bar{\cal J}$, instead of that for ${\cal J}$.

\no
One may wonder how one could have constructed \eqn{jjdre} without resorting to previous work, in particular \eqn{apam1},
but using the action \eqn{sklinf} as the only input. The way to proceed in this direction
is  firstly to  realize that beyond the free field limit an operator mixing occurs giving rise to a non-diagonal anomalous dimension matrix.
Diagonalizing it will determine the operators with well defined anomalous dimension. The first step in this procedure is to determine the set of operators which in principle may mix. This consists by the operators having the same
classical anomalous dimension in the free theory limit. For the case at hand, the set of operators with classical dimension one is
\be
{\cal O}^a_{\pm,n} = (f^n)^{ab}\del_\pm\phi^b\ ,  \qq n=0,1,2, \dots\ .
\ee
In  the free theory limit  these have vanishing two-point functions among themselves for different
$n$'s, so that they constitute a diagonal basis.
Subsequently, one computes the same two point functions to a given order in the $1/k$-expansion which generically, will no longer be diagonal. They can be subsequently diagonalized to the given order in the $1/k$-expansion and their anomalous
dimensions can be read off. Obviously there will be no mixing between opposite chirality operators since the theory
is Lorentz invariant. The operators \eqn{jjdre} with $h_1$ and $h_2$ given by \eqn{g12} provide by construction the first operators in this hierarchy having a well defined anomalous dimension.
In order to compute the corrected version of the partner operator that appears in  the mixing, i.e. $f\del_\pm\phi$, $f^2\del_\pm \phi$ etc, we need
to evaluate more expressions like the ones in \eqn{233} which will not do in this paper.

\subsection{Anomalous dimensions of primary fields}

Consider the field $D^{ab}$ as defined in \eqn{jjd}.
Then to ${\cal O}(1/k)$ we do not need any insertions coming from the exponential of the action.  As a result one obtains
\be
\begin{split}
\langle D^{ac}(x_1,\bar x_1) D^{b c}(x_2,\bar x_2)\rangle
& =  \d_{ab} +{1\ov k}{1-\l\ov 1+\l} \langle f^{ac}(x_1,\bar x_1) f^{bc}(x_2,\bar x_2)\rangle
\\
& = \d_{ab}\Big(1 +{c_G\ov k}{1-\l\ov 1+\l} \ln{\e^2\ov |x_{12}|^2}\Big)\ ,
\end{split}
\ee
from which the anomalous dimension of $D^{ab}$ can be read
\be
\label{gammD}
\boxed{
\g_D = {c_G\ov k}{1-\l\ov 1+\l}} \ .
\ee
In fact we may also compute the anomalous dimension of the group element $g(x,\bar x)$ which is a primary field in some
irreducible representation $R$ with unitary matrices $t_a$. As before using \eqn{ggrouu} and the rescaling \eqn{xphii}
we compute that
\be
\begin{split}
\langle g^{ik}(x_1,\bar x_1) g^{-1}_{kj}(x_2,\bar x_2)\rangle
& =  \d_{ij} +{1\ov k}{1-\l\ov 1+\l} (t_a)_{ik} (t_b)_{kj} \langle \phi^a(x_1,\bar x_1) \phi^b(x_2,\bar x_2)\rangle
\\
& = \d_{ij}\Big(1 +{c_R\ov k}{1-\l\ov 1+\l} \ln{\e^2\ov |x_{12}|^2}\Big)\ ,
\end{split}
\ee
where $(t_at_a)_{ij}=c_R \d_{ij}$, with $c_R$ being the quadratic Casimir in the representation $R$. From the last equation 
we deduce that
\be
\boxed{
\g_g = {c_R\ov k}{1-\l\ov 1+\l}
} \ ,
\ee
in agreement with previous work (see eq. (3.26) of \cite{Georgiou:2016iom} after setting $c_{R'}=c_R$ since the representations matrices are the same for the left and the right transformations). Note that specializing to the adjoint representation we obtain the anomalous
dimension \eqn{gammD} for the composite field of $g$, that is the dimension of $D_{ab}$ which is not
something that one should have necessarily expected. It is not clear that this equality relation
will persist beyond the leading order in the $1/k$-expansion.
Note also that, unlike the current operators, the primary field operators we have considered here are not $\l$-dressed. 
i.e. they retain their bare CFT expressions.

\section{The $\b$-function}

In this section, we will derive the expression for the running of the parameter $\l$ under the renormalization group flow,
by using the background field method.
In the context of $\l$-deformations this method was first used in
\cite{Appadu:2015nfa}, albeit in a first order formulation of the equations of motion.
In order to obtain the $\b$-function for the coupling $\l$ it is enough to look at the renormalization
of the cubic coupling in \eqn{sklinf}. We will see that unlike the case of the correlation functions of the previous section, in this case all three couplings up to fifth order in the action \eqn{sklinf} will be involved, as well. We will also show that that the renormalization of the quartic coupling in \eqn{sklinf} will give rise to the same $\b$-function, as it should be, in a non-trivial manner. In that case the sixth order coupling is necessary as well.

To proceed we need the equations of motion for the fields $\phi^a$. Varying the action up to
terms of  ${\cal O}(1/k^{2})$ we obtain that\footnote{A direct variation of the action \eqn{sklinf}
gives an equation of the form
$\displaystyle \Big(\d_{ab}+{g_4\ov k} f^2_{ab}+ {g_6\ov k^2} f^4_{ab}\Big)\del_+\del_-\phi^b +\dots =0$.
The following expression is obtained after multiplying with the inverse of the matrix prefactor and keeping terms up to the specified order.}
\be
\label{eqs345}
\begin{split}
& \del_+ \del_- \phi^a -{3g_3\ov 2\sqrt{k}} f_{abc}\del_+\phi^b \del_-\phi^c
-{g_4\ov k}(f_{abd}f_{dc}+f_{acd}f_{db})\del_+\phi^b\del_-\phi^c
\\
&
\qq +{1\ov 2k^{3/2}}\Big( 3 g_3 g_4 f^2_{ad} f_{dbc}-g_5 \big(\del_a f^3_{bc}+ {\rm cyclic\ in}\  a,b,c\big) \Big)\del_+\phi^b\del_-\phi^c
\\
&\qq + {1\ov k^2} \Big(g_4^2 f^2_{ae}(f_{ebd} f_{dc} + f_{ecd} f_{db})
-{g_6\ov 2} (\del_a f^4_{bc} - \del_b f^4_{ac} - \del_c f^4_{ab})  \Big) \del_+\phi^b\del_-\phi^c = 0\ .
\end{split}
\ee
Next we compute the fluctuation $\d \phi^a$ of this equation, around a classical solution which we we will still denote by $\phi^a$. We will cast them in the form
\be
\label{dhatt}
\hat D^{ab}\d\phi^b = 0\ ,
\ee
where the operator $\hat D$ is second order in the worldsheet derivatives.
We will present its explicit expression after the Euclidean analytic continuation and in momentum space. In the conventions of \cite{Georgiou:2018hpd}, we replace $(\del_+,\del_-) $ by $ \ha (\bar p,p)\equiv (p_+,p_-)$. Then after dividing by $p_+p_-$ the operator $\hat D$ takes the form
\be
\begin{split}
& \hat D^{ab} = \d^{ab} + \hat F^{ab} \ ,
\\
& \hat F^{ab} = {g_3\ov \sqrt{k}}\, F^{'ab}_1 + {g_4\ov k}\, F_2^{ab} + {g_4\ov k} F_2^{'ab} + {1\ov k^{3/2}}\, F_3^{ab}  + {1\ov k^{3/2}}\, F_3^{'ab}
+ {1\ov k^2}\, F_4^{ab}\ ,
\end{split}
\ee
where we have used the following definitions
\be
\label{Fsp}
\begin{split}
& F^{'ab}_1 = {3\ov 2} f_{abc} \Big( {\del_+\phi^c \ov p_+} - {\del_-\phi^c \ov p_-}\Big)\ ,
\\
&
 F^{ab}_2 = {B^{abcd}\ov p_+p_-}\, \del_+\phi^c \del_-\phi^d\ ,
\qq F^{'ab}_2 = B^{adbc} \Big( {\del_+\phi^c \ov p_+} + {\del_-\phi^c \ov p_-}\Big)\phi^d\ ,
\\
&
F_3^{ab} = {C^{abcd}\ov p_+p_-}\, \del_+\phi^c \del_-\phi^d\ ,
\qq F^{'ab}_3 = -C^{abc} \Big( {\del_+\phi^c \ov p_+} - {\del_-\phi^c \ov p_-}\Big)\ ,
\\
&
F_4^{ab} = {D^{abcd}\ov p_+p_-}\, \del_+\phi^c \del_-\phi^d\ .
\end{split}
\ee
Note the Lorentz non-invariant terms which we have indicated by a prime.
Nevertheless, as we will see they will eventually combine into a Lorentz invariant result.
For that matter we did not keep the term, which would have been denoted as $F_4'$ , arising
from varying the derivatives $\del_+\phi^b\del_- \phi^c$ in the last line in \eqn{eqs345}.
Such a term should have to combine with $F_1'$ in order to give a Lorentz invariant result.
But then apparently the result would be of order ${\cal O}(1/k^{5/2})$ and as such 
would contribute to the fifth coupling interaction term.
Moreover, we have found it convenient to define the following tensors
\be
\begin{split}
 & B^{abcd} = f_{acm} f_{mbd} + f_{adm} f_{mbc} = B^{abdc}\ ,
\\
&
C^{abcd} = {3\ov 2} g_3 g_4 \del_b f^2_{am} f_{mcd} - \ha g_5 \big(\del_a\del_b f^3_{cd}+{\rm cyclic\ in}\ a,c,d \big)=- C^{abdc} \ ,
\\
&
C^{abc}= {3\ov 2} g_3 g_4 f^2_{ad} f_{dbc} - {g_5\ov 2} \big(\del_a f^3_{bc}
+ {\rm cyclic\ in}\  a,b,c\big)= -C^{acb}\ ,
\\
& D^{abcd} = \del_b\Big(g^2_4 f^2_{ae}\big(f_{ecg}f_{gd}+f_{edg}f_{gc}\big)-{g_6\ov 2}
\del_b\big(\del_a f^4_{cd}- \del_c f^4_{ad} - \del_d f^4_{ac}\big)\Big)= D^{abdc}\ .
\end{split}
\ee
Integrating out the fluctuations, gives the effective Lagrangian of our model
\be
\label{lefff}
-\cL_{\rm eff} = \cL^{(0)}_{k,\l} + \int^\m {d^2 p\ov (2\pi)^2} \ln (\det \hat D)^{-1/2}\ ,
\ee
where $\cL^{(0)}_{k,\l}$  is the Lagrangian of the action \eqn{sklinf}.
This integral is logarithmically divergent with respect to the UV mass scale
$\m$. The divergence is isolated by performing the large momentum expansion of the integrand
and keeping terms proportional to $\displaystyle {1\ov |p|^2}$, where $|p|^2=p\bar p$.
We use the identity
\be\label{ident}
\ln (\det \hat D)  = \Tr\hat F -\ha \Tr\hat F^2 + \cdots  \ .
\ee
The traces in \eqn{ident} can be easily calculated to give
\be
\label{tracees}
\begin{split}
&
\Tr \hat F = {g_4\ov k} \Tr F_2 + {1\ov k^{3/2}} \Tr F_3 + {1\ov k^2} \Tr F_4 \ ,
\\
&
\Tr \hat F^2 = {g_3^2\ov k}\, \Tr F^{'2}_1 + 2{g_3 g_4\ov k^{3/2}}\, \Tr(F'_1 F'_2)
+ {g_4^2\ov k^2}\, \Tr F^{'2}_2  + 2{g_3\ov k^2}\, \Tr(F'_1 F'_3)
   \ ,
\end{split}
\ee
where in the first line we have not included terms having a single power of $1/p_+$ or
$1/p_-$, since these terms  will give zero when the angular part of the integration  in \eqn{lefff} is performed.

\no
In the following we examine the  renormalization of  the cubic and the quartic interaction vertices and show  that they both
give rise to the same $\b$-function for $\l$, thanks to  the specific dependence of the couplings $g_i$, $i=3,\dots , 6$ on the single parameter $\l$.
We will see that in the process wave-function renormalization, as well as field redefinitions will be needed.
This is expected from the work of \cite{Itsios:2014lca}. In that work the gravitational approach in computing
the $\beta$-function using the full action  \eqn{sdddg} was employed and coordinate reparamaterizations were indeed
needed. These in a field theoretical language would correspond to wave function and field redefinitions.

\subsection{Renormalization of the cubic vertex}

For the cubic vertex we only need to consider terms up to order ${\cal O}(1/k^{3/2})$.
We start by explicitly computing the relevant traces in \eqn{tracees}
\be
\begin{split}
& \Tr F_2 = {2 c_G\ov p_+ p_-}\,  \del_+\phi^a \del_-\phi^a \ ,
\\
& \Tr F_3 = {c_G\ov 2 p_+ p_-}  (3 g_3 g_4 + 5 g_5)\, f_{ab} \del_+\phi^a \del_-\phi^b\ ,
\\
&
\Tr F^{'2}_1 = {9 c_G\ov 2 p_+ p_-}\, \del_+\phi^a \del_-\phi^a\ ,
\\
&
\Tr (F'_1 F'_2) = {9 c_G\ov 2 p_+ p_-}\, f_{ab} \del_+\phi^a \del_-\phi^b \ .
\end{split}
\ee
The next step is to use polar coordinates, i.e. $p=re^{i\phi}$, $\bar p=re^{-i\phi}$, in which
the integration measure reads  $d^2 p = r dr d\phi$ and subsequently evaluate the effective action.
Since we need terms up to ${\cal O}(1/k^{3/2})$ we truncate the action keeping only up to the cubic interaction term.
Equation \eqn{lefff} then straightforwardly gives (for clarity, we return back to the Lorentzian regime)
\be
\label{hk2ui1}
\begin{split}
& S_{\rm eff}= {1\ov 2\pi} \int d^2\s\, \bigg[ \Big[1-{c_G\ov k}\Big(2 g_4 -{9\ov 4} g_3^2\Big)\ln \m^2\Big]\del_+ \phi^a \del_- \phi^a
\\
& \qq\qquad + {1\ov \sqrt{k}}
\Big[g_3-{c_G\ov k}\Big({5\ov 2} g_5 -3 g_3 g_4\Big)\ln \m^2\Big] f_{ab}\del_+ \phi^a \del_- \phi^b \bigg]+\cdots\ .
\end{split}
\ee
The wavefunction renormalization
\be
\phi^a = Z^{1/2} \hat \phi^a\ , \qq Z =1 + {c_G\ov k}\Big(2 g_4 -{9\ov 4} g_3^2\Big)\ln \m^2\ .
\label{dfgg1}
\ee
puts the kinetic term into a canonical form and \eqn{hk2ui1} becomes
\be
\label{hk2ui2}
\begin{split}
& S_{\rm eff}= {1\ov 2\pi} \int d^2\s\, \bigg[ \del_+ \hat \phi^a \del_- \hat \phi^a
\\
& \qq\quad + {1\ov \sqrt{k}}
\Big[g_3-{c_G\ov k}\Big({5\ov 2} g_5 -6 g_3 g_4 +{27\ov 8}g_3^3\Big)\ln \m^2\Big]
\hat  f_{ab}\del_+ \hat \phi^a \del_- \hat \phi^b \bigg] +\cdots\ .
\end{split}
\ee
We demand that the action \eqn{hk2ui1} is $\m$-independent, i.e. $\del_{\ln \m^2} \cL_{\rm eff}=0$.
For $k\gg 1$ this derivative acts only on the coupling constant $g_3(\l)$.
Then, we obtain that
\be
\b^\l g'_3 = {c_G\ov 8k} \big(20 g_5 -48 g_3 g_4 +27g_3^3\big)\ ,\qq
\b^\l \equiv {\m\ov 2}\, {d\l\ov d\m} \ ,
 \ee
from which using the explicit expressions \eqn{g34} the $\b$-function for the coupling $\l$
\be
\label{systrg}
\boxed{
\b^\l  = -{c_G\ov 2k} {\l^2\ov (1+\l)^2} }\ ,
\ee
follows. This is precisely the expression firstly computed in \cite{Kutasov:1989dt} and  \cite{Gerganov:2000mt} with 
CFT methods and in \cite{Itsios:2014lca} with gravitational methods. 

\subsection{Renormalization of the quartic vertex}

In this case we should consider the terms of  order ${\cal O}(1/k^2)$ as well.
Performing the corresponding traces in \eqn{tracees} is much more difficult. Nevertheless, we have performed this task for
the case where the group $G$ is $SU(2)$. Then, in our normalizations
\be
f_{abc}=\sqrt{2} \e_{abc}\ ,\qq c_G = 4\ .
\ee
Then
\be
f^2_{ab}= 2(\phi_a \phi_b- \d_{ab} { \phi}^2)\ ,\qq  f^3_{ab}=- 2\sqrt{2}\e_{abc}\phi_c {\phi}^2\ ,
\ee
where ${ \phi}^2=\phi^a \phi^a$.
In addition, we compute the $C^{abc}$ and $B^{abcd}$ tensors which take the form 
\be
\begin{split}
& C^{abc}=  3 \sqrt{2} (g_5-g_3 g_4)\e_{abc}\phi^2
\\
&\qq + \sqrt{2}(3 g_3g_4 + 2 g_5) \e_{abcd} \phi^a\phi^d + 2 \sqrt{2} g_5(\e_{abd}\phi^c - \e_{acd}\phi^b)\phi^d\ ,
\\
& B^{abcd} = 4 \d_{ab} \d_{cd} -2 \d_{ad} \d_{bc} - 2\d_{ac} \d_{bd} \ .
\end{split}
\ee
Subsequently, we calculate the additional traces in \eqn{tracees} to be
\be
\begin{split}
& \Tr F_4 = 12(2g_4^2+g_6)\,  \phi_a\phi_b \del_+\phi^a \del_- \phi^b
 - 4 (2 g_4^2+ 11 g_6)\,  \phi^2 \del_+\phi^a \del_- \phi^a\ ,
 \\
 &
\Tr F^{'2}_2  = 48\,  \phi_a\phi_b \del_+\phi^a \del_- \phi^b -32\,   \phi^2 \del_+\phi^a \del_- \phi^a\ ,
\\
&
\Tr(F'_1 F'_3)  = 18 g_3g_4\,   \phi_a\phi_b \del_+\phi^a \del_- \phi^b
+ 2 (9g_3 g_4-30 g_5)\,   \phi^2 \del_+\phi^a \del_- \phi^a\ .
\end{split}
\ee
As a result, the quartic interaction term of the action becomes
\be
\label{hk2ui3}
\begin{split}
& S_{\rm eff}^{(4)}= {1\ov 2\pi} \int d^2\s\, {2\ov k}\bigg[
\Big[ g_4-  {1\ov k}\Big(6 g_6 -9 g_3^2 g_4\Big)\ln \m^2\Big] \phi^a\phi^b\del_+ \phi^a \del_- \phi^b
\\
&\qq\quad -  \Big[ g_4-  {1\ov k}\Big(22 g_6 +9 g_3^2 g_4 - 4 g_4^2 -30 g_3 g_5\Big)\ln \m^2\Big] \phi^2\del_+ \phi^a \del_- \phi^a\bigg] \ .
\end{split}
\ee
However, one should not forget to take into account the wave-function renormalization \eqn{dfgg1} (with $c_G=4$). Doing so, the  quartic interaction term of the action becomes
\be
\label{hk2ui4}
\begin{split}
& S_{\rm eff}^{(4)}= {1\ov 2\pi} \int d^2\s\, {2\ov k}\bigg[
\Big[ g_4-  {1\ov k}\Big(6 g_6 +9 g_3^2 g_4-16 g_4^2\Big)\ln \m^2\Big]
\hat\phi^a\hat\phi^b\del_+ \hat\phi^a \del_- \hat\phi^b
\\
&\qq\quad -  \Big[ g_4-  {1\ov k}\Big(22 g_6 +27 g_3^2 g_4 - 20 g_4^2 -30 g_3 g_5\Big)\ln \m^2\Big] \hat\phi^2\del_+ \hat\phi^a \del_- \hat\phi^a\bigg] \ .
\end{split}
\ee
Clearly, demanding that this action \eqn{hk2ui1} is $\m$-independent would lead to two different expressions for the
$\b$-function for $\l$ and in fact none of  them coincides with \eqn{systrg}. In order to make the two compatible one needs to perform
a field redefinition. Realizing that $f_{ab}\phi_b=0$, an appropriate ansatz can only be of the form
\be
\label{hfdhh}
\hat \phi^a = \hat Z^{1/2} \tilde \phi^a\ ,\qq \hat Z^{1/2} = 1+ {\hat C\ov k^2}  \tilde \phi^2 \ln \m^2 \ ,
\ee
where $\hat C$ is a couping dependent constant. Applying this field redefinition to \eqn{hk2ui4} we get
\be
\label{hk2ui5}
\begin{split}
& S_{\rm eff}^{(4)}= {1\ov 2\pi} \int d^2\s\, {2\ov k}\bigg[
\Big[ g_4-  {1\ov k}\Big(6 g_6 +9 g_3^2 g_4-16 g_4^2-2 \hat C\Big)\ln \m^2\Big]
\tilde\phi^a\tilde\phi^b\del_+ \tilde\phi^a \del_- \tilde\phi^b
\\
&\qq\quad -  \Big[ g_4-  {1\ov k}\Big(22 g_6 +27 g_3^2 g_4 - 20 g_4^2 -30 g_3 g_5+\hat C\Big)\ln \m^2\Big] \tilde\phi^2\del_+ \tilde\phi^a \del_- \tilde \phi^a\bigg] \ ,
\end{split}
\ee
where the terms involving $\hat C$ arise from the contribution of the quadratic term in \eqn{hk2ui2} due to the field
redefinition in \eqn{hfdhh} above.  Demanding now that $\del_{\ln \m^2} \cL_{\rm eff}=0$ leads to two differential equations for $g_4$ whose compatibility requires that
\be
\hat C= {4\ov 3}g_4^2 +10 g_3 g_5 -6 g_3^2 g_4 -{16\ov 3}g_6\ .
\ee
Then we obtain that
\be
\label{fhji1}
\b^\l g'_4 = {1\ov 3 k} \big(63 g_3^2 g_4 -56 g_4^2 -60 g_3 g_5 + 50 g_6)\ .
 \ee
Substituting in the last equation the values for the couplings one  gets the same $\b$-function as \eqn{systrg}  (wth $c_G=4$).
For general groups the factor $1/3$ on the right hand side of \eqn{fhji1} is expected to be replaced by $c_G/12$, the reason being that for large $k$ the perturbative expansion is in terms of the ratio $c_G/k$.

\no
It should be clear that we may keep the analysis general order by order in perturbation theory and derive
a system of first order differential equations for the $g_i$'s, $i=3,4,\dots $.
Demanding that all couplings depend only on a single parameter $\l$ with the above $\b$-function, this procedure
generates all couplings to arbitrarily high order once the first three ones  are known. 
For instance, \eqn{fhji1} with the $\b$-function \eqn{systrg} generates $g_6(\l)$ and so on and so forth.
The summed up action should of course be the $\l$-deformed action \eqn{djkg11}.
Note that, the RG structure of the theory is such that the flow is between the UV CFT point and a gapped IR theory when the coupling becomes strong. In our free field approach the strong
coupling regime can be and indeed is achieved since the coupling constants $g_i$ are not independent, but all have a very specific dependence on a single coupling $\l$. Moreover, 
they all have singularities at $\l=\pm 1$.

\section{Discussion and future directions}

In the  present work we have established a systematic method for performing various computations 
in a wide class of integrable $\s$-models that go under the name of $\l$-deformed models.
These include computations of two- and three-point functions of current and primary filed operators, as well as the $\b$-functions as {\it exact} functions of the deformation parameter $\l$. 
As an explicit example, we have chosen to perform our calculations in the simplest case of the single $\l$-deformed model \cite{Sfetsos:2013wia}.
All of our results are in complete agreement with the expressions obtained 
earlier in\cite{Georgiou:2016iom,Georgiou:2015nka} and are obtained after considerably easier computational efforts.  

The main idea of our method is that instead of using perturbation theory around the conformal point it is more effective to use it around the free point.
Practically, we have set up a large $k$ perturbative expansion of the effective action of the model  around the free field point corresponding to the identity group element $g=\mathbb{I}$. 
The resulting action defines a two-dimensional quantum field theory with a canonical kinetic term and infinite number of interaction terms involving successively more and more fields. For our purposes, we have kept all field interaction terms up to sixth order. An additional ingredient in our approach was the free field expansion of the $\l$-dressed operators which we
have also established.
One of the virtues of our method is that it incorporates automatically  all the dependence in $\l$, that is all-loop effects, since these are already built in at the action level and as such they inherit the non-perturbative symmetry in coupling space \eqn{invv}. 
A second advantage of our approach is that it can be considered as the basis for systematically performing the perturbative expansion of the $\b$-function in powers of $1/ k$. The reason is that, one may focus at the renormalization of a single vertex,
i.e. the cubic being the simplest, instead of seeking the renornalization of the entire $\sigma$-model effective action. 
A similar comment concerns the higher, in $1/k$, corrections of the anomalous dimensions.

There are several directions towards this line of research can be extended. First recall that the original $\l$-deformed model
\cite{Sfetsos:2013wia}
was formulated for general deformation matrix $\l_{ab}$, but correlation functions and operator anomalous dimensions 
are not known beyond the case of diagonal  matrix we have studied also here. 
The present method is particularly suitable to handle the arbitrary matrix case as well. 
In addition, it would be 
interesting to reproduce for the case of left/right asymmetric models \cite{Georgiou:2017jfi}, and by using our method, the exact results for the correlation functions of currents and/or primaries, as well as that for the $\b$-function that were obtained earlier in \cite{Georgiou:2016zyo} by the use of conformal perturbation theory.  Secondly, it would be important to apply our method to study the quantum properties of the wide class of coupled integrable models constructed recently in \cite{Georgiou:2018hpd} and \cite{Georgiou:2018gpe}. We note that in these models although the running of the couplings is known we lack exact results involving correlation functions of currents and primary fields. 
Finally, as mentioned above, our set up is ideal for calculating the subleading ${1/ k^2}$ corrections to the $\beta$-functions, anomalous dimensions and correlatiors both for the single and the multiply $\l$-deformed models. Such a calculation will provide a non-trivial check of the non-perturbative symmetry of the model \cite{Kutasov:1989aw} beyond the 
leading order in the $1/ k$-expansion. It will also clarify certain issues concerning the $1/ k$ corrections to the effective action. Another direction would
be to find other non-trivial classical solutions of the $\l$-deformed theories which would be the analogue of the uniton solutions found in the case of the PCM. 
It would be certainly interesting to apply our method and expand around these solutions instead of expanding around the identity.

\no
In addition to $\l$-deformation corresponding to integrable deformations of WZW currents algebra CFTs, there is the similar construction in which the deformed CFT is based on a symmetric coset \cite{Sfetsos:2013wia,Hollowood:2014rla} or the $AdS_5 \times S^5$ Superstring \cite{Hollowood:2014qma}. It will be of importance to extend our present consideration in these
cases as well.  

\no
Next, we comment on the relation of the free field expansion \eqn{sklinf} to 
the fact that the action \eqn{djkg11} from which it originates is integrable. 
To motivate the argument note that if all the coupling coefficients of the terms $f^n_{ab} \del_+\phi^a \del_-\phi^b$ in the free
field expansion \eqn{sklinf} were set to unity then this would correspond to the free field expansion of the non-Abelian T-dual of PCM for the group $G$ \cite{Giveon:1993ai}, which is an integrable $\s$-model (see, for instance, appendix D of \cite{Sfetsos:2013wia}). 
Placing arbitrary coefficients in front of the interaction terms and demanding integrability 
would place stringent conditions on them in the form of recursive relations. One solution to these  would be the sequence of coefficients, all depending on a single 
coupling $\l$, arising from the free field 
expansion of  the $\l$-deformed action \eqn{djkg11}, the first few terms being  given by \eqn{sklinf} and \eqn{g34}. Another integrable case corresponds to the pseudo-chiral $\s$-model
\cite{Nappi:1979ig} 
in which all couplings vanish except for $g_3$. Note that, as a mathematical truncation, this is consistent with the renormalization group flow equations as well. 
This approach could be useful in trying to find other solutions to the aforementioned 
recursive relations that could depend on more parameters. The corresponding 
integrable $\s$-models could in fact go beyond the $\l$-deformed models in the sense that it is
far from obvious that they can be obtained by a gauging procedure. 

\no
  Finally, we believe that, as technique, the free field expansion, with appropriate modifications will be useful in determining 
  anomalous dimensions and quantum properties in general of operators in the so-called  $\eta$-deformations \cite{Klimcik:2002zj,Klimcik:2008eq,Klimcik:2014,Delduc:2013fga,Delduc:2013qra,Arutyunov:2013ega} using their also
  their relation to
  $\l$-deformations \cite{Vicedo:2015pna,Hoare:2015gda,Sfetsos:2015nya}.
Similar comment applies for the integrable coupled $\sigma$-models of \cite{Delduc:2018hty,Delduc:2019bcl}.

\subsection*{Acknowledgments}

The work of G.G. on this project has received funding from the Hellenic Foundation for Research and Innovation
(HFRI) and the General Secretariat for Research and Technology (GSRT), under grant
agreement No 15425. \\
K.S. would like to thank the CERN Theoretical Physics Department for
hospitality and financial support during part of this research.

\appendix

\section{Anomalous dimensions: Computational details}

In this appendix we present the details of the computation of the most laborious integrals the are required for
the evaluation of the anomalous dimensions and of the three-point correlation functions of the currents of the currents .

\subsection{Current anomalous dimension}

In this appendix we give  the details of the computation of the integral $I_2^{ab}$
defined in \eqn{I123} since they are rather involved compared to $I_1^{ab}$ and $I_3^{ab}$.
 Picking up and performing the first Wick contraction with
$\bar J_2^{b_2}(\z_2)$, this splits as
\be
I_2^{ab}=I_{2,1}^{ab} + I_{2,2}^{ab} + I_{2,3}^{ab}\ ,
\ee
where the various terms are given by
\be
\begin{split}
& I_{2,1}^{ab}= C \int d^2z_1\ \langle J^b(x_2) J^{a_1}(z_1) f^{a_1 a_2}(z_1,\z_1) \bar J^{a_2}(\z_1) J^{b_1}(x_1) f^{b_1a}(x_1,\bar x_1) \rangle
\\
&\qq\qq\qq + [(x_1,a)\leftrightarrow (x_2,b)]\ ,
\end{split}
\ee
by that
\be
I_{2,2}^{ab}= f^{a_1a_2 b_2}\int {d^2 z_1 d^2 z_2\ov \z_{12}}\ \langle J^a(x_1) J^b(x_2) J^{a_1}(z_1) \bar J^{a_2}(\z_1) J^{b_1}(z_2) f^{b_1b_2}(z_2,\z_2) \rangle\
\ee
and by that
\be
I_{2,3}^{ab}= -\int {d^2 z_1 d^2 z_2\ov \z_{12}^2}\ \langle J^a(x_1) J^b(x_2) J^{a_1}(z_1) f^{a_1 a_2}(z_1,\z_1) J^{b_1}(z_2) f^{b_1 a_2}(z_2,\z_2)\rangle\ .
\ee
We observe that $I_{2,1}^{ab}$ equals the integral $I_3^{ab}$ in \eqn{i3ab} times the factor $C=\pi$. Therefore,
using \eqn{i3ab} we immediately obtain that
\be
\label{I21ab}
I_{2,1}^{ab}=  - 6\pi^2\, {c_G \d^{ab}\ov x_{12}^2}\,
\ln{\e^2\ov |x_{12}|^2}\ .
\ee
Turning to the second integral and using for the Wick contraction $\bar J^{a_2}(\z_1)$,
we obtain
\be
\begin{split}
&
I_{2,2}^{ab}= C f^{a_1a b_2}\int  {d^2 z_2\ov \bar x_1-\z_2}\ \langle  J^b(x_2) J^{a_1}(x_1) J^{b_1}(z_2)
f^{b_1 b_2}(z_2,\z_2) \rangle + [(x_1,a)\leftrightarrow (x_2,b)]
\\
&\qq\qq -  f^{a_1a_2 b_2} f^{a_2b_1b_2}\int  {d^2 z_1 d^2 z_2\ov \z_{12}^2}\,
\langle J^a(x_1) J^b(x_2) J^{a_1}(z_1) J^{b_1}(z_2)\rangle
\\
& \qq\qq = C f^{a_1 a b_2} f^{b_2 a_1 b} \int {d^2 z_2\ov (\z_2-\bar x_1)}{1\ov (z_2-x_1)(z_2-x_2)} \Big({1\ov z_2-x_2}-
{1\ov z_2-x_1}\Big)
\\
&\qq\qq\qq + [(x_1,a)\leftrightarrow (x_2,b)]
\\
&\qq\qq + 2 c_G \d^{ab} \int {d^2 z_1 d^2 z_2\ov \z_{12}^2 (z_1-x_1)^2 (z_2-x_2)^2} \ ,
\end{split}
\ee
where we have included only integrals that potentially contribute and we have also used \eqn{EGG1}.
As a result, one obtains that
\be
\begin{split}
&I_{2,2}^{ab}=  c_G C\d^{ab} \int  d^2 z_2\ \Big({1\ov (z_2-x_2)^2(z_2-x_1)(\z-\bar x_1)} -
{1\ov (z_2-x_1)^2(z_2-x_2)(\z-\bar x_1)}\Big)
\\
& \qq\qq + [(x_1,a)\leftrightarrow (x_2,b)]
\\
& \qq\qq + 2\pi^2 c_G\d^{ab}\int d^2z_2\ {\d^{(2)}(z_2-x_1)\ov (z_2-x_2)^2}\ ,
\end{split}
\ee
where we have used the last integral in \eqn{B1}. Using also the second integral in \eqn{B2} and setting $C=\pi$ we obtain
\be
\label{I22ab}
I_{2,2}^{ab}=  - 4\pi^2\, {c_G \d^{ab}\ov x_{12}^2}\,
\ln{\e^2\ov |x_{12}|^2}\ .
\ee
Finally, the last integral to be computed is
\be
\label{i23}
\begin{split}
I_{2,3}^{ab} & =  2 \int {d^2 z_1 d^2 z_2\ov \z_{12}^2 (z_1-x_1)^2 }\langle J^b(x_2) f^{aa_2}(z_1,\z_1) J^{b_1}(z_2)
f^{b_1 a_2}(z_2,\z_2)\rangle
\\
& + 2 f^{aa_1 a_2}   \int {d^2 z_1 d^2 z_2\ov \z_{12}^2 (x_1-z_1) }
\langle J^b(x_2) J^{a_1}(z_1) J^{b_1}(z_2) f^{b_1a_2}(z_2,\z_2)\rangle\ .
\end{split}
\ee
The first line of \eqn{i23} gives
\be
\label{shshss}
\begin{split}
& 2 c_G \d^{ab} \int {d^2 z_1 d^2 z_2\ov \z_{12}^2 (z_1-x_1)^2 }\, \Big({\ln |z_{12}|^2\ov (z_2-x_2)^2 }-
{1\ov z_{12}  (z_2-x_2)}\Big)
\\
&= 2\pi c_G \d^{ab}\Bigg( \int {d^2 z_2 \ov (z_2-x_2)^2(z_2-x_1) (\z_2-\bar x_1)} +
\int {d^2 z_1 \ov (z_1-x_1)^2(z_1-x_2) (\z_1-\bar x_2)} \Bigg)
\\
& =-(2\pi^2+2\pi^2){c_G \d^{ab}\ov x_{12}^2}\,
\ln{\e^2\ov |x_{12}|^2}\ ,
\end{split}
\ee
where we have used \eqn{B3} and the second and third integrals of \eqn{B2}.
We now move to the second line in \eqn{i23} which gives
\be
\label{kjsahdfkjl}
-2 c_G \d^{ab} \int {d^2z_1 d^2z_2\ov \z_{12}^2  z_{21} } {1\ov (z_1-x_1)(z_2-x_2)}
\Big({1\ov z_2-x_2}-{1\ov z_{21}}\Big)\ .
\ee
Note that the first integral in \eqn{kjsahdfkjl} is identical to the second integral in the first line in \eqn{shshss} (after exchanging the indices $1$ and $2$ in both the $z$'s and the $x$'s) giving a
contribution of $ \displaystyle - 2\pi^2\, {c_G \d^{ab}\ov x_{12}^2}\,
\ln{\e^2\ov |x_{12}|^2}$.

\no
The second integral in \eqn{kjsahdfkjl} is
\ba
\label{last}
&& 
2c_G \d^{ab}  \int {d^2z_2\ov  z_2-x_2} \int {d^2z_1\ov \z_{12}^2  z_{12}^2  (z_1-x_1)} = 
2 c_G \d^{ab} \int {d^2z_2\ov  z_2-x_2}
 \int d^2 z_1 \Big[\del_{\z_1} {-1\ov \z_{12} z_{12}^2 (z_1-x_1)}
\nonumber
 \\
 && +\pi {\d^{(2)}(z_1-x_1)\ov \z_{12} z_{12}^2} \Big]=
 2 c_G \d^{ab} \int {d^2z_2\ov  z_2-x_2}\Big[{i\ov 2}\oint {dz_1 \ov \z_{12} z_{12}^2
 (z_1-x_1)} + {\pi\ov (\bar x_1 -\z_2)(x_1-z_2)^2}\Big]
\nonumber \\
 &&\qq =
  -2\pi^2{c_G \d^{ab}\ov x_{12}^2}\, \ln{\e^2\ov |x_{12}|^2}\ .
\ea
This result arises from the second integration over $z_2$ which appears in the second line of \eqn{last}.
To evaluate this second integral we have employed the first relation in \eqn{B2}. 
The line integral of \eqn{last} is infinite and can be absorbed by anappropriate counterterm (see below).
Therefore we finally obtain
\be
\label{I23ab}
I_{2,3}^{ab}=  - 8\pi^2 {c_G \d^{ab}\ov x_{12}^2}\,
\ln{\e^2\ov |x_{12}|^2}\ . 
\ee
Hence, adding up \eqn{I21ab}, \eqn{I22ab} and \eqn{I23ab} gives \eqn{i2ab}.

\section{Some useful integrals}

In our computation we encounter the following divergent integrals which we have regulated by introducing a small distance cut-off $\e$
\be\label{B1}
\begin{split}
&
\int \frac{d^2z}{(x_1-z)(\z-\bar{x}_1)}=\pi\, \text{ln}\, \epsilon^2 \ ,
\\
&
\int \frac{d^2z}{(x_1-z)(\z-\bar{x}_2)}=\pi\, \text{ln}|x_{12}|^2 \ ,
\\
&
\int \frac{d^2z}{(x_1-z)^2(\z-\bar{x}_2)}=-\frac{\pi}{x_{12}} \ ,
\\
&
\int\frac{d^2z}{(x_1-z)(\z-\bar{x}_2)^2}=-\frac{\pi}{\bar{x}_{12}} \ ,
\\
&
\int\frac{d^2z}{(x_1-z)^2(\z-\bar{x}_2)^2}=\pi^2\, \delta^{(2)}(x_{12}) \ ,
\end{split}
\ee
as well as, 
\be\label{B2}
\begin{split}
&
\int\frac{d^2z}{(z-x_1)^2(z-x_2)(\z-\bar{x}_1)}=\frac{\pi}{x_{12}^2}\, \text{ln}\frac{\e^2}{|x_{12}|^2} \ ,
\\
&
\int\frac{d^2z}{(z-x_1)^2(z-x_2)(\z-\bar{x}_2)}=-\frac{\pi}{x_{12}^2}\, \text{ln}\frac{\epsilon^2}{|x_{12}|^2}-\frac{\pi}{x_{12}^2}  \ ,
\\
&
\int \frac{d^2z}{(z-x_1)^2(\z-\bar{x}_1)(\z-\bar{x}_2)}=\frac{\pi}{|x_{12}|^2}\ ,
\\
&
\int\frac{d^2z}{(z-x_1)(z-x_2)^2(\z-\bar{x}_3)^2}
=-\frac{\pi}{x_{12}^2}\frac{\bar{x}_{12}}{\bar{x}_{13}\bar{x}_{23}}
-\pi^2{\d^{(2)}(x_{23})\ov x_{12}}\ .
\end{split}
\ee
Finally, we will also need the integral
\be\label{B3}
\int d^2z\frac{\ln|z-x_1|^2}{(z-x_2)^2(\z-\bar{x}_1)^2}=\frac{\pi}{|x_{12}|^2} \ .
\ee


\begin{thebibliography}{1}


\bibitem{Faddeev:1994zg} 
  L.~D.~Faddeev and G.~P.~Korchemsky,
  {\it High-energy QCD as a completely integrable model},
  Phys.\ Lett.\ B {\bf 342}, 311 (1995)
  \href{http://arxiv.org/abs/hep-th/9404173}{hep-th/9404173}.
 
  
  
\bibitem{Minahan:2002ve} 
  J.~A.~Minahan and K.~Zarembo,
  {\it The Bethe ansatz for N=4 superYang-Mills},\hfill\break
  JHEP {\bf 0303}, 013 (2003)
   \href{http://arxiv.org/abs/hep-th/0212208}{hep-th/0212208}.
 
  
  
\bibitem{Bena:2003wd} 
  I.~Bena, J.~Polchinski and R.~Roiban,
  {\it Hidden symmetries of the AdS(5) x $S^5$ superstring},
  Phys.\ Rev.\ D {\bf 69}, 046002 (2004)
 \href{http://arxiv.org/abs/hep-th/0305116}{hep-th/0305116}.
  %

 \bibitem{Sfetsos:2013wia}
  K.~Sfetsos, {\it Integrable interpolations: From exact CFTs to non-Abelian T-duals},\hfill\break
  Nucl. Phys. {\bf B880} (2014) 225, \href{http://arxiv.org/abs/arXiv:1312.4560}{arXiv:1312.4560 [hep-th]}.
 


\bibitem{Georgiou:2016urf}
  G.~Georgiou and K.~Sfetsos,
  {\it A new class of integrable deformations of CFTs},\\
   JHEP {\bf 1703} (2017) 083,
  \href{https://arxiv.org/abs/1612.05012}{arXiv:1612.05012 [hep-th]}.

  \bibitem{Georgiou:2017jfi}
  G.~Georgiou and K.~Sfetsos,
  {\it Integrable flows between exact CFTs},\\
  JHEP {\bf 1711}  (2017) 078,
   \href{https://arxiv.org/abs/1707.05149}{arXiv:1707.05149 [hep-th]}.
   
   \bibitem{Sfetsos:2017sep}
  K.~Sfetsos and K.~Siampos,
  {\it Integrable deformations of the $G_{k_1} \times G_{k_2}/G_{k_1+k_2}$ coset CFTs},
  Nucl. Phys. {\bf B927} (2018) 124,
  \href{https://arxiv.org/abs/1710.02515}{arXiv:1710.02515  [hep-th]}.

\bibitem{Georgiou:2018hpd}
  G.~Georgiou and K.~Sfetsos,
  {\it Novel all loop actions of interacting CFTs: Construction, integrability and RG flows},
  Nucl. Phys. {\bf B937} (2018) 371,\href{https://arxiv.org/abs/1809.03522}{arXiv:1809.03522 [hep-th]]}.

\bibitem{Georgiou:2018gpe}
  G.~Georgiou and K.~Sfetsos,
  {\it The most general $\lambda$-deformation of CFTs and integrability},
    JHEP {\bf 1903} (2019) 094,
  \href{https://arxiv.org/abs/1812.04033} {arXiv:1812.04033 [hep-th]}.

\bibitem{Driezen:2019ykp}
  S.~Driezen, A.~Sevrin and D.~C.~Thompson,
  {\it Integrable asymmetric $\lambda$-deformations},
  JHEP {\bf 1904}, 094 (2019)
  \href{https://arxiv.org/abs/1902.04142}{arXiv:1902.04142 [hep-th]}.

   \bibitem{Sfetsos:2014lla}
  K.~Sfetsos and K.~Siampos,
 {\it The anisotropic $\lambda$-deformed SU(2) model is integrable},
  Phys. Lett. {\bf B743} (2015) 160,
  \href{http://arxiv.org/abs/1412.5181}{arXiv:1412.5181 [hep-th]}.
  
  
   \bibitem{Georgiou:2015nka}
  G.~Georgiou, K.~Sfetsos and K.~Siampos,
  {\it All-loop anomalous dimensions in integrable $\lambda$-deformed $\sigma$-models},
  Nucl.\ Phys.\  {\bf B901} (2015) 40,
  \href{http://arxiv.org/abs/1509.02946}{arXiv:1509.02946 [hep-th].}
  
\bibitem{Georgiou:2016iom}
  G.~Georgiou, K.~Sfetsos and K.~Siampos,
  {\it All-loop correlators of integrable $\l$-deformed $\s$-models},
  Nucl.  Phys. {\bf B909} (2016) 360,
  \href{http://arxiv.org/abs/arXiv:1604.08212}{1604.08212 [hep-th].}

  \bibitem{Georgiou:2016zyo}
  G.~Georgiou, K.~Sfetsos and K.~Siampos,
  {\it $\lambda$-deformations of left-right asymmetric CFTs}, Nucl. Phys. {\bf B914} (2017) 623,
\href{https://arxiv.org/abs/1610.05314}{arXiv:1610.05314 [hep-th]}.

\bibitem{Georgiou:2017oly}
  G.~Georgiou, K.~Sfetsos and K.~Siampos,
  {\it Double and cyclic $\lambda$-deformations and their canonical equivalents},
  Phys. Lett. {\bf B771}  (2017) 576,
   \href{https://arxiv.org/abs/1704.07834}{arXiv:1704.07834 [hep-th]}.
   
    \bibitem{Itsios:2014lca}
  G.~Itsios, K.~Sfetsos and K.~Siampos,
  {\it The all-loop non-Abelian Thirring model and its RG flow},
  Phys.\ Lett.\  {\bf B733} (2014) 265,
  \href{http://arxiv.org/abs/1404.3748}{arXiv:1404.3748 [hep-th].}

 
     \bibitem{Georgiou:2017aei}
  G.~Georgiou, E.~Sagkrioti, K.~Sfetsos and K.~Siampos,
  {\it Quantum aspects of doubly deformed CFTs},
Nucl. Phys. {\bf B919} (2017) 504,
 \href{https://arxiv.org/abs/1703.00462}
  {arXiv:1703.00462 [hep-th]}.
  
  \bibitem{Witten:1983ar}
  E.~Witten,
  {\it Nonabelian Bosonization in Two-Dimensions},\hfill\break
  \href{https://link.springer.com/article/10.1007\%2FBF01215276}{Commun. Math. Phys.\  {\bf 92} (1984) 455.}


  
  \bibitem{Balog:1993es}
  J.~Balog, P.~Forgacs, Z.~Horvath and L.~Palla,
  {\it A New family of $SU(2)$ symmetric integrable sigma models,}
  Phys. Lett. {\bf B324} (1994) 403,
  \href{http://arxiv.org/abs/hep-th/9307030}{hep-th/9307030.}

    

\bibitem{Georgiou:2019jcf} 
  G.~Georgiou, P.~Panopoulos, E.~Sagkrioti and K.~Sfetsos,
  {\it Exact results from the geometry of couplings and the effective action},\hfill\break 
   Nucl. Phys. {\bf B948} (2019) 114779,
   \href{https://arxiv.org/abs/1906.00984}{arXiv:1906.00984 [hep-th]}.


\bibitem{Hoare:2018jim} 
  B.~Hoare, N.~Levine and A.~A.~Tseytlin,
  {\it On the massless tree-level S-matrix in 2d sigma models},
  J. Phys. {\bf A52}, no. 14, 144005 (2019),
  \href{https://arxiv.org/abs/1812.02549}
  {arXiv:1812.02549 [hep-th]}.
  
  
 \bibitem{Kutasov:1989aw}
  D.~Kutasov, {\it Duality Off the Critical Point in Two-dimensional Systems With Nonabelian Symmetries},
\href{http://www.sciencedirect.com/science/article/pii/0370269389913257}{Phys. Lett. {\bf B233} (1989) 369}.

\bibitem{Pol}
J.~Polchinski, {\it Superstring theory, Vol. 1}, Cambridge University Press 1998.


   \bibitem{Kutasov:1989dt}
  D.~Kutasov,
  {\it String Theory and the Nonabelian Thirring Model},\\
 \href{http://www.sciencedirect.com/science/article/pii/0370269389912859}{Phys. Lett. {\bf B227} (1989) 68}.

  \bibitem{Gerganov:2000mt}
  B.~Gerganov, A.~LeClair and M.~Moriconi,
  {\it On the beta function for anisotropic current interactions in 2-D},
  Phys. Rev. Lett. {\bf 86} (2001) 4753,
 \href{http://arxiv.org/abs/hep-th/0011189}{hep-th/0011189}.

 
  \bibitem{Appadu:2015nfa}
  C. Appadu and T.J. Hollowood,
  {\it Beta function of $k$ deformed ${\text AdS}_{5} \times S^5$ string theory},
  JHEP {\bf 1511} (2015) 095,
  \href{http://arxiv.org/abs/arXiv:1507.05420}{arXiv:1507.05420 [hep-th].}


 
\bibitem{Hollowood:2014rla}
  T.J.~Hollowood, J.L.~Miramontes and D.M.~Schmidtt,
 {\it Integrable Deformations of Strings on Symmetric Spaces},
  JHEP {\bf 1411} (2014) 009,
  \href{http://arxiv.org/abs/1407.2840}{arXiv:1407.2840 [hep-th]}.



\bibitem{Hollowood:2014qma}
  T.J.~Hollowood, J.L.~Miramontes and D.~Schmidtt,
{\it An Integrable Deformation of the $AdS_5 \times S^5$ Superstring},
J. Phys. {\bf A47} (2014) 49,  495402,
 \href{http://arxiv.org/abs/1409.1538}{arXiv:1409.1538 [hep-th]}.

\bibitem{Giveon:1993ai}
  A.~Giveon and M.~Rocek,
  {\it On non-Abelian duality}
  Nucl. Phys. {\bf B421} (1994) 173,\break
    \href{http://arxiv.org/abs/hep-th/9308154}{hep-th/9308154}.


  \bibitem{Nappi:1979ig}
  C.R. Nappi,
  {\it Some Properties of an Analog of the Nonlinear $\sigma$-Model},\\
 \href{ http://journals.aps.org/prd/abstract/10.1103/PhysRevD.21.418}
  {Phys. Rev. {\bf D21} (1980) 418}.

 \bibitem{Klimcik:2002zj}
C. Klim\v c\'\i k,
  {\it YB sigma models and dS/AdS T-duality},\hfill\break
  JHEP {\bf 0212} (2002) 051,
\href{http://arxiv.org/abs/hep-th/0210095}{hep-th/0210095}.

\bibitem{Klimcik:2008eq}
  C. Klim\v c\'\i k,
  {\it On integrability of the YB sigma-model},\hfill\break
  J. Math. Phys. {\bf 50} (2009) 043508,
  \href{http://arxiv.org/abs/0802.3518}{arXiv:0802.3518 [hep-th]}.

 \bibitem{Klimcik:2014}
  C. Klim\v c\'\i k,
  {\it Integrability of the bi-Yang--Baxter sigma-model},
  Letters in Mathematical Physics {\bf 104} (2014) 1095,
    \href{http://arxiv.org/abs/1402.2105}{arXiv:1402.2105 [math-ph]}.

   \bibitem{Delduc:2013fga}
  F.~Delduc, M.~Magro and B.~Vicedo,
{\it On classical $q$-deformations of integrable sigma-models},
  JHEP {\bf 1311} (2013) 192,
   \href{http://arxiv.org/abs/1308.3581}{arXiv:1308.3581 [hep-th]}.

\bibitem{Delduc:2013qra}
  F.~Delduc, M.~Magro and B.~Vicedo,
{\it An integrable deformation of the $AdS_5 \times S^5$ superstring action},
  Phys. Rev. Lett. {\bf 112}, 051601,
     \href{http://arxiv.org/abs/1309.5850}{arXiv:1309.5850 [hep-th].}

\bibitem{Arutyunov:2013ega}
  G.~Arutyunov, R.~Borsato and S.~Frolov,
  {\it S-matrix for strings on $\eta$-deformed $AdS_{5} \times S^5$},
  JHEP {\bf 1404} (2014) 002,
 \href{http://arxiv.org/abs/1312.3542}{arXiv:1312.3542 [hep-th].}
  
  \bibitem{Vicedo:2015pna}
  B.~Vicedo,
  {\it Deformed integrable $\sigma$-models, classical $R$-matrices and classical exchange algebra on Drinfel'd doubles},
  \hfill\break
  J. Phys. A: Math. Theor. {\bf 48} (2015) 355203,
 \href{http://arxiv.org/abs/1504.06303}{arXiv:1504.06303 [hep-th].}

\bibitem{Hoare:2015gda}
  B.~Hoare and A.A.~Tseytlin,
  {\it On integrable deformations of superstring sigma models related to $AdS_n \times S^n$ supercosets},\hfill\break
  {Nucl. Phys. {\bf B897} (2015) 448},
  \href{http://arxiv.org/abs/1504.07213}{arXiv:1504.07213 [hep-th].}



 \bibitem{Sfetsos:2015nya}
  K.~Sfetsos, K.~Siampos and D.C.~Thompson,
 {\it Generalised integrable $\lambda$- and $\eta$-deformations and their relation}, \hfill\break
  Nucl. Phys. {\bf B899} (2015) 489,
  \href{http://arxiv.org/abs/1506.05784}{arXiv:1506.05784 [hep-th]}.
  
\bibitem{Delduc:2018hty}
  F.~Delduc, S.~Lacroix, M.~Magro and B.~Vicedo, {\it Integrable coupled sigma-models},
   Phys. Rev. Lett. {\bf 122} (2019) no.4, 041601,
  \href{https://arxiv.org/abs/1811.12316}{ arXiv:1811.12316 [hep-th]}.

\bibitem{Delduc:2019bcl}
  F.~Delduc, S.~Lacroix, M.~Magro and B.~Vicedo,
  {\it Assembling integrable $\sigma$-models as affine Gaudin models},
  JHEP {\bf 1906} (2019) 017,
   \href{https://arxiv.org/abs/1903.00368}{  arXiv:1903.00368 [hep-th]}.





  \end{thebibliography}
\end{document}